\documentclass[10pt]{article}
\def \BE {\begin{equation}}
\def \EE {\end{equation}}
\def \BEA {\begin{eqnarray}}
\def \EEA {\end{eqnarray}}
\def \CR {\nonumber \\}

\def \e {{\epsilon}}

\begin{document}
\title{Probability densities and 
preservation of randomness in wave turbulence}
\author{Yeontaek Choi$^*$, Yuri V. Lvov$^\dagger$ and Sergey Nazarenko$^*$
\\
$^*$
Mathematics Institute, The University of Warwick,  Coventry, CV4 7AL, UK \\
$^\dagger$ Department of Mathematical Sciences, Rensselaer
Polytechnic Institute, Troy, NY 12180 }

\maketitle

\begin{abstract}
Turbulence closure for the weakly nonlinear stochastic waves requires,
besides weak nonlinearity, randomness in both the phases and the
amplitudes of the Fourier modes. This randomness, once present
initially, must remain over the nonlinear evolution time.
Finding out to what extent is this true is the main goal of the
present Letter.   For this analysis we derive an evolution equation for
the full probability density function (PDF)  of the wave field.
We will show that, for any statistics of the amplitudes, phases tend
to stay random if they were  random initially.  If in
addition the initial amplitudes are independent variables they will
remain independent in a coarse-grained sense, i.e.  when considered in
small subsets which are much less than the total set of modes.
\end{abstract}

\section{Introduction}

%
The theory of stochastic wavefields in weakly nonlinear dispersive
media has a long and exciting history which started in 1929 when
Peierls derived his kinetic equation for phonons in solids
\cite{peierls}. Applications of these ideas
appeared in the physics of the ocean and atmosphere
\cite{Zakfil,hasselman,BS,BN,Newell68,ZakharovPRL,LT}, laboratory and
astrophysical plasmas \cite{GS,davidson,lvovzakh}, Bose condensates
and nonlinear optics \cite{DNPZ}, anharmonic crystals
\cite{peierls,bp,prig}.  Any attempt to give a fair historical review
would be doomed in such a short letter and we refer an interested
reader for further references to the book \cite{ZLF} and a more recent
review \cite{NNB}.  The common name that has arisen for all these
approaches is Wave Turbulence (WT).

WT closure requires, besides weak nonlinearity, randomness in both the
phases and the amplitudes of the Fourier modes. Namely, all the phases
and all the amplitudes must be statistically independent of each
other, in some sense, and the phases must be uniformly
distributed. Such an approach was recently formulated in \cite{ln,cln} as
a generalization of the 
Random Phase
Approximation (RPA) much loved by the physicists  which, in its traditional form, ignores the
amplitude randomness \cite{ZLF}. We even kept the same acronym RPA but
now read it as ``Random Phases and Amplitudes''.  Below, in section
\ref{RPA}, we define explicitly what we mean by RPA. RPA does not
fix the shape of the probability densities of the individual mode
amplitudes and, therefore, it allows one to consider wavefields with
non-decaying correlations which is helpful because such long
correlations tend to arise naturally in WT systems.  In \cite{ln}, we
used RPA to describe the arbitrary-order moments of the wave
amplitude, and in \cite{cln} we extended this approach to describing
the one-mode probability density function (PDF) and considered
solutions for this PDF corresponding to intermittency. In these works,
however, RPA was assumed (but not proven) to hold over the nonlinear
time.

Such a proof is the main goal of the present paper.  We shall consider
initial fields of the RPA type, and we will prove that the 
RPA properties are preserved (i.e. no phase or amplitude
correlations are generated with accuracy sufficient for the WT
 closure)  over the nonlinear evolution
time.  In order to do this we shall derive an evolution equation for
the full multi-mode PDF which will turn out to be the
Zaslavski-Sagdeev (ZS) equation \cite{zs} (a WT cousin of the
Brout-Prigogine equation for anharmonic crystals \cite{bp,prig}).  We
will show that, for any statistics of the amplitudes, phases tend to
stay random if they were so initially.  If, in addition, the initial
amplitudes are independent variables they will remain independent in a
coarse-grained sense, i.e.  when considered in small subsets which are
much smaller than the total set of modes.

The original paper by ZS \cite{zs} was also devoted to the study of
the applicability of the WT closure and, therefore, it is appropriate here
to mention in which way our approach is different.  First, ZS consider
the nonlinear interaction arising from the potential energy only
(i.e. the interaction Hamiltonian involves coordinates but not
momenta). This restriction leaves out the capillary water waves,
Alfven, internal and Rossby waves, as well as many other interesting
WT systems. In our work we remove this restriction by considering the
most general three-wave Hamiltonian equation (\ref{Interaction}) and
we show that the multi-mode PDF still obeys the ZS equation in this
case.  Secondly, ZS studied the phase statistics only, whereas our
work considers both the phases and the amplitudes because the
amplitude statistics is as important for the RPA closure as the phase
statistics.  Thirdly, ZS presented an argument that the nonlinear
frequency correction removes the need for the initial phase
randomness, whereas we only state the preservation of the initial phase
randomness. However, the ZS criterion for
phase randomization was obtained from a rather non-rigorous (although
highly intuitive) physical argument whereas our results follow from a
systematic asymptotic expansion outlined in this Letter and the
details of which will be published in a more extended paper
\cite{cln1}.

The validation of the RPA properties gives this technique the status of
a well-justified approach which, due to the simplicity of its
premises, is a winning tool for the future theory of non-Gaussianity
of WT, its intermittency and interactions with coherent structures.

\section{Statistical setup. }
Let us consider a wavefield $a({\bf x}, t)$ in a periodic cube of with
side $L$ and let the Fourier transform of this field be $a_l(t)$ where
index $l {\in } {\cal Z}^d$ marks the mode with wavenumber $k_l = 2
\pi l /L$ on the grid in the $d$-dimensional Fourier space.  For
simplicity let us assume that there is a maximum wavenumber $k_{max}$
(fixed e.g. by dissipation) so that no modes with wavenumbers greater
than this maximum value can be excited.  In this case, the total
number of modes is $N = (k_{max} / \pi L)^d$. Correspondingly, index
$l$ will only take values in a finite box, $l \in {\cal B}_N \subset
{\cal Z}^d$ which is centered at 0 and all sides of which are equal to
$k_{max} / \pi L = N^{1/3}$.  To consider homogeneous turbulence, the
large box limit $N \to \infty $ will have to be taken.
\footnote{ It is easy to extend the analysis to the infinite Fourier
space, $k_{max} = \infty$.  In this case, the full joint PDF would
still have to be defined as a $N \to \infty$ limit of an $N$-particle
PDF, but this limit would have to be taken in such a way that both
$k_{max}$ and the density of the Fourier modes tend to infinity
simultaneously.}

Let us write the complex $a_l$ as $a_l =A_l \psi_l $ where $A_l$ is a
real positive amplitude and $\psi_l $ is a phase factor which takes
values on ${\cal S}^{1} $, a unit circle centered at zero in the
complex plane. Let us define the $N$-particle joint PDF ${\cal
P}^{(N)}$ as the probability for the wave intensities $A_l^2 $ to be
in the range $(s_l, s_l +d s_l)$ and for the phase factors $\psi_l$ to
be on the unit-circle segment between $\xi_l$ and $\xi_l + d\xi_l$ for
all $l \in {\cal B}_N$.  In terms of this PDF, taking the averages
will involve integration over all the real positive $s_l$'s and along
all the complex unit circles of all $\xi_l$'s,
\BEA \langle f\{A^2, \psi \} \rangle
= \left(
 \prod_{ l {\cal 2 B}_N }  \int_{{\cal R}^{+} } ds_l \oint_{{\cal S}^{1} }
|d \xi_l| \right) \;  {\cal P}^{(N)} \{s, \xi \}
f\{s, \xi \} \label{pdfn} \EEA
where the notation $f\{A^2,\psi\}$ means that $f$ depends on all $A_l^2$'s
and all $\psi_l $'s in the set $\{A_l^2, \psi_l; l {\cal 2 B}_N \}$
(similarly, $\{s, \xi \}$ means $\{s_l, \psi_l; l \in {\cal B}_N \}$,
etc). The full PDF that contains the complete statistical information
about the wavefield $a({\bf x}, t)$ in the infinite $x$-space can be
understood as a large-box limit
$${\cal P} \{ s_k, \xi_k \}  =  \lim_{N \to \infty}
{\cal P}^{(N)} \{s, \xi \},
$$
i.e. it is a functional acting on the continuous functions of the
wavenumber, $s_k$ and $\xi_k$.  In the the large box limit there is a
path-integral version of (\ref{pdfn}),
\BE \langle f\{A^2, \psi \} \rangle =
 \int {\cal D}s \oint
|{\cal D} \xi| \;  {\cal P} \{s, \xi \}  f\{s, \xi \}
\label{mean-path} \EE
The full PDF defined above involves all $N$ modes (for either finite
$N$ or in the $N \to \infty$ limit). By integrating out all the
arguments except for chosen few, one can have reduced statistical
distributions. For example, by integrating over all the angles and
over all but $M$ amplitudes,we have an ``$M$-particle'' amplitude PDF,
\BE
{\cal P}_{j_1, j_2, \dots , j_M} = \left(
\prod_{ l \ne j_1, j_2, \dots , j_M }  \int_{{\cal R}^{+} } ds_l 
\prod_{ m {\cal 2 B}_N } 
\oint_{{\cal S}^{1} }
|d \xi_m| \; \right) {\cal P}^{(N)} \{s, \xi \},
\EE
which depends only on the $M$ amplitudes marked by labels
$j_1, j_2, \dots , j_M  {\cal 2 B}_N$. 

Statistical derivations are greatly facilitated by the introduction
of a generating functional
\BEA
Z^{(N)} \{\lambda, \mu \} =  
{ 1 \over (2 \pi)^{N}} \langle
\prod_{l \in {\cal B}_N }   \; e^{\lambda_l A_l^2} \psi_l^{\mu_l}
\rangle , \label{Z}
\EEA
where $\{\lambda, \mu \} \equiv \{\lambda_l, \mu_l ; l \in {\cal
B}_N\}$ is a set of parameters, $\lambda_l {\cal 2 R}$ and $\mu_l
{\cal 2 Z}$.
\BE 
{\cal P}^{(N)}  \{s, \xi \} = 
 {1 \over (2 \pi)^{N}}  \sum_{\{\mu \}}   \langle
\prod_{l \in {\cal B}_N } \delta (s_l - A_l^2) \, \psi_l^{\mu_l}
\xi_l^{-\mu_l} \rangle = \hat {\cal L}_\lambda^{-1}
 \sum_{\{\mu \}}
 \left( Z^{(N)} \{\lambda, \mu\}
\, \prod_{l \in {\cal B}_N } \xi_l^{-\mu_l} \right) \label{jointpdf}
\EE
where $\{\mu \} \equiv \{ \mu_l \in {\cal Z}; {l \in {\cal B}_N } \}$
is a set of indices enumerating the angular harmonics and $\hat {\cal
L}_\lambda^{-1}$ stands for the inverse Laplace transform with respect to all
$\lambda_l$.

\subsection{Definition of an essentially RPA field\label{RPA}}

A {\em pure} RPA fields can be defined as one in which all the
phases and amplitudes of the Fourier modes make a set of $2N$
statistically independent variables and in which all phase factors
$\psi$ are uniformly distributed on their respective unit circles. In
such pure form RPA never survives except for in the un-interesting state
of complete thermodynamic equilibrium. However, WT closure only
requires an approximate RPA which holds up to certain order in
small $\epsilon$ and $1/N$ and only in a coarse-grained sense, i.e.
for the reduced $M$-particle objects with $M \ll N$.  Below we give a
relaxed definition of an (essentially) RPA property which, on one
hand, is sufficient for the WT closure and, on the other hand, is
preserved over the nonlinear time.

{\bf Definition:} We will say that the field $a$ is of an {\em
essentially RPA} type if:

\begin{enumerate}

\item The phase factors are statistically independent and uniformly
distributed variables up to $O(\e^2)$ corrections, i.e.
\BE
{\cal P}^{(N)} \{s, \xi \}  = {1 \over (2 \pi)^{N} } {\cal P}^{(N,a)} \{s \} 
 \; [1 +O(\e^2)],
\EE
where 
\BE
 {\cal P}^{(N,a)} \{s \} = 
\left(
\prod_{ l {\cal 2 B}_N } 
\oint_{{\cal S}^{1} }
|d \xi_l| \; \right) {\cal P}^{(N)} \{s, \xi \},
\EE
is the $N$-particle {\em amplitude} PDF.  In terms of the generating
functional
\BE Z^{(N)} \{\lambda, \mu \}  =  Z^{(N,a)}  \{\lambda \}
\, \prod_{l \in {\cal B}_N } \delta(\mu_l) \; [1 +O(\e^2)],
\label{z-rpa} \EE
where
\BE 
 Z^{(N,a)}  \{\lambda \}
=\langle 
\prod_{l \in {\cal B}_N } 
e^{\lambda_l A_l^2} \rangle
= Z^{(N)} \{\lambda, \mu\}|_{\mu=0}
\EE
is an $N$-particle generating function for the amplitude statistics.

\item
The amplitude variables are independent in a {\em coarse-grained}
sense, i.e. for each $M \ll N$ modes the $M$-particle amplitude PDF is
equal to the product of the one-particle PDF's up to $O(M/N)$ and
$o(\e^2)$ corrections,
\BE
{\cal P}^{(M,a)}_{j_1, j_2, \dots , j_M} = 
 P^{(a)}_{j_1}  P^{(a)}_{j_2} \dots  P^{(a)}_{j_M} \; [1 +
O(M/N) + O(\e^2)].
\label{second}
\EE
\end{enumerate}
As a first step in validating the RPA property we will have to prove
that the generating functional remains of the form (\ref{z-rpa}) over the
nonlinear time provided it has this form at $t=0$.

\section{  Weak-nonlinearity expansion.}

Consider weakly nonlinear dispersive waves in a periodic box with a
dispersion relation $\omega_k$ which allow three-wave
interactions. Example of such systems include surface capillary
waves~\cite{Zakfil,{ZakharovPRL}}, Rossby waves~\cite{bnaz} and
internal waves in the ocean~\cite{LT}.  In Fourier space, we have the
following Hamiltonian equations,
\BEA i \, \dot a_l &=& \epsilon \sum_{m,n=1}^\infty \left(
V^l_{mn} a_{m} a_{n}e^{i\omega_{mn}^l t} \, \delta^l_{m+n}
 + 2 \bar{V}^{m}_{ln} \bar a_{n}
a_{m} e^{-i\omega^m_{ln}t } \, \delta^m_{l+n}\right),
\label{Interaction} \EEA
where $a_l=a(k_l)$ is the complex wave amplitude in the interaction
representation, $k_l = 2 \pi l/L $ is the wavevector, $L $ is the box
side length,
$\omega^l_{mn}\equiv\omega_{k_l}-\omega_{k_m}-\omega_{k_m}$,
$\omega_l=\omega_{k_l}$ is the wave frequency, $\epsilon \ll 1$ is a
formal nonlinearity parameter. Here, the interaction coefficient
$V^l_{mn}$ is obviously symmetric with respect to $m$ and $n$ but we
do not assume any further symmetries.\footnote{Some additional
symmetries involving permutations of the upper and lower indices
arise, e.g., in solids due to the fact that nonlinearity is purely due
to the potential energy which is a function of the displacement but
not the rate of the displacement. Refs. \cite{bp,prig,zs} imposed such
symmetries which immediately rule out the capillary, internal and
other waves in fluids for which such properties do not
hold. Additional symmetries also arise if the action variable is a
Fourier transform of a real quantity, e.g., in the Rossby waves
\cite{bnaz}.}

In order to filter out fast oscillations at the wave period, let us
seek for the solution at time $T$ such that $2 \pi / \omega \ll T \ll
1/\omega \epsilon^2$.  The second condition ensures that $T$ is a lot
less than the nonlinear evolution time.  Now let us use a perturbation
expansion in small $\epsilon$,
\BE a_l(T)=a_l^{(0)}+\epsilon a_l^{(1)}+\epsilon^2 a_l^{(2)}.
\label{Expansion} \EE
Substituting this expansion in (\ref{Interaction}) we get in the
zeroth order
$ a_l^{(0)}(T)=a_l(0)\label{definitionofa} $,
i.e. the zeroth order term is time independent. This corresponds to
the fact that in the interaction representation, wave amplitudes are
constant in the linear approximation.  For simplicity, we will write
$a^{(0)}_l(0)= a_l$, understanding that a quantity is taken at $T=0$
if its time argument is not mentioned explicitly.  The first order is
given by
\BEA a^{(1)}_l (T) = -i \sum_{m,n=1}^\infty \left(   V^l_{mn} a_m
a_n \Delta^l_{mn} \delta^l_{m+n}
 + 2
\bar{V}^m_{ln}a_m\bar{a}_n \bar\Delta^m_{ln}\delta^m_{l+n}
\right), \label{FirstIterate} \EEA
where $ \Delta^l_{mn}=\int_0^T e^{i\omega^l_{mn}t}d t =
({e^{i\omega^l_{mn}T}-1})/{i \omega^l_{mn}}. \label{NewellsDelta}
$
Iterating one more time we get 
\BEA a_l^{(2)} (T)  &=& \sum_{m,n, \mu, \nu}^\infty \left[ 2
V^l_{mn} \left( -V^m_{\mu \nu}a_n a_\mu a_\nu E[\omega^l_{n \mu
\nu},\omega^l_{mn}] \delta^m_{\mu + \nu} -2 \bar V^\mu_{m \nu}a_n
a_\mu \bar a_\nu \bar E[\omega^{l \nu}_{n
\mu},\omega^l_{mn}]\delta^\mu_{m + \nu}\right) \delta^l_{m+n}
\right.\CR && \left. + 2 \bar V^m_{ln}
 \left(-V^m_{\mu \nu}\bar a_n a_\mu a_\nu E[\omega^{ln}_{\mu \nu},-\omega^m_{ln}]
\delta^m_{\mu + \nu} - 2 \bar V^\mu_{m \nu}\bar a_n a_\mu \bar
a_\nu E[-\omega^\mu_{n \nu l},-\omega^m_{l n}]  \delta^\mu_{m +
\nu} \right) \delta^m_{l+ n} \right. \CR && \left. + 2 \bar
V^m_{ln} \left( \bar V^n_{\mu \nu}a_m \bar a_\mu  \bar a_\nu
\delta^n_{\mu + \nu} E[-\omega^m_{l\nu\mu},-\omega^m_{ln}] + 2
V^\mu_{n \nu}a_m \bar a_\mu  a_\nu E[\omega^{\mu l}_{\nu m},
-\omega^m_{ln}]\delta^\mu_{n + \nu}\right)\delta^m_{l+n}
\right],\CR\label{SecondIterate} \EEA

\noindent where we introduced
$E(x,y)=\int_0^T \Delta(x-y)e^{i y t} d t .$
%

\section{Evolution of the Generating Functional and Multi-particle PDF}

%
Let us first derive an evolution equation for the generating
functional $Z\{\lambda, \mu\}$ exploiting the separation of the linear
and nonlinear time scales. \footnote{ Hereafter we omit superscript
${(N)}$ in the $N$-particle objects if it does not lead to a
confusion.}  To do this, we have to calculate $Z$ at the intermediate
time $t=T$ based on its value at $t=0$. The derivation, although
standard for WT, is quite lengthy and will have to be published in a
longer paper.  Here, we will only outline the main steps and give the
result. First, we need to substitute the $\epsilon$-expansion of $a$
from (\ref{Expansion}) into the expressions $e^{\lambda_j |a_j|^2}$
and $\psi_j^{\mu_j} ={1 \over 2 } (\ln {a_j \over \bar a_j})^{\mu_j}
$. Second, the phase averaging should be done.  Note that, because, we
assume that initial phase factors are independent at $t=0$ with
required accuracy, we can do such phase averaging independently of the
amplitude averaging (which we do not do yet).  Thirdly, we take $N \to
\infty$ limit followed by $T \sim 1/\epsilon \to \infty$ (this order
of the limits is essential!).  Taking into account that
$\lim\limits_{T\to\infty}E(0,x)= T (\pi \delta(x)+iP(\frac{1}{x}))$,
and $\lim\limits_{T\to\infty}|\Delta(x)|^2=2\pi T\delta(x)$ and,
replacing $(Z(T) -Z(0))/T$ by $\dot Z$ (because the nonlinear time
$\sim 1/\epsilon^2 \gg T)$ we have
\BEA
\dot Z	&=&
4 \pi \e^2
 \int \big\{ (\lambda_{j}+\lambda_{j}^2 {\delta \over \delta \lambda_{j}})
\left[|V_{mn}^{j}|^2 \delta(\omega_{mn}^{j}) \delta_{m+n}^{j} +
2 |V_{jn}^{m}|^2 \delta(\omega_{jn}^{m}) \delta_{j+n}^{m}
\right]
{\delta^2 Z\over \delta \lambda_{m} \delta \lambda_{n}} 
\nonumber \\
&& + 2
\lambda_j
\left[ - |V_{mn}^j|^2 \delta(\omega_{mn}^j)\delta_{m+n}^j
{\delta \over \delta \lambda_{n}}
+|V_{jn}^m|^2 \delta(\omega_{jn}^m)\delta_{j+n}^m 
\left(
{\delta \over \delta \lambda_{m}}
- {\delta \over \delta \lambda_{n}} \right)
\right]  
{\delta Z \over \delta \lambda_{j}}
\nonumber \\
&& + 
2 \lambda_j\lambda_m
\left[ 
-2 |V_{mn}^j|^2 \delta_{m+n}^j \delta(\omega_{mn}^j)
+|V_{jm}^n|^2 \delta_{j+m}^n \delta(\omega_{jm}^n) 
\right]
{\delta^3 Z \over \delta \lambda_{j} \delta \lambda_{n}
\delta \lambda_{m}} \big\}\, dk_j dk_m dk_n.
\label{Zequat}
\EEA
Here variational derivatives appeared instead of partial derivatives
because of the $N\to\infty$ limit.  This expression is valid up to the
$[1+O(\e^2)]$ factor.  Equation (\ref{Zequat}) does not contain $\mu$
dependence which means that that these variables separate from
$\lambda$'s and the solution is a purely-amplitude $Z$ times an
arbitrary function of $\mu$'s which is going to be stationary in time.
The latter corresponds to preservation of the initial $\Pi
\delta(\mu_l)$ dependence by equation (\ref{Zequat}) which means that
no angular harmonics of the PDF higher than zeroth will be excited. In
the other words, all the phases will remain statistically independent
and uniformly distributed on $S^1$ with the accuracy of the equation
(\ref{Zequat}) integrated over the nonlinear time $1/\e^2$, i.e. with
the $O(\e^2)$ accuracy. This proves the first of the ``essential RPA''
properties. In fact, this result was already obtained before in
\cite{bp} for a narrower class of 3-wave systems (see footnote 2).
Note that we still have not used any assumption about the statistics
of $A$'s and, therefore, (\ref{Zequat}) could be used in future for
studying systems with random phases but correlated amplitudes.

Taking the inverse Laplace transform of (\ref{Zequat}) we have the
following equation for the PDF,
\BE
\dot {\cal P} = - \int {\delta F_j \over \delta s_j} \, dk_j,
\label{zse}
\EE
where $F_j$ is a flux of probability in the space of the amplitude $s_j$,
\BEA
-{F_j \over
4 \pi \e^2 s_j} 
&=&
\int
\big\{ 
(|V_{mn}^{j}|^2 \delta(\omega_{mn}^{j}) \delta_{m+n}^{j} +
2 |V_{jm}^{n}|^2 \delta(\omega_{jm}^{n}) \delta_{j+m}^{n}
)
s_n s_m {\delta {\cal P} \over \delta s_j}
\nonumber \\
&&
+2 {\cal P} (
|V_{jm}^{n}|^2 \delta(\omega_{jm}^{n}) \delta_{j+m}^{n}
- |V_{mn}^{j}|^2 \delta(\omega_{mn}^{j}) \delta_{m+n}^{j} 
)s_m
\nonumber \\
&&
+2
(|V_{jm}^{n}|^2 \delta(\omega_{jm}^{n}) \delta_{j+m}^{n}
-2|V_{mn}^{j}|^2 \delta(\omega_{mn}^{j}) \delta_{m+n}^{j} )
s_n s_m
{\delta {\cal P} \over \delta s_m}
 \big\} \, dk_m dk_n
\label{flux}
\EEA
This equation is identical to the Zaslavski-Sagdeev (ZS) \cite{zs}
equation (Brout-Prigogine in the physics of crystals context
\cite{bp,prig}).  Note that ZS equation was originally derived in
\cite{zs} for a much narrower class of systems, see footnote 2,
whereas the result above indicates that it is also valid in the most
general case of 3-wave systems.  Here we should again emphasize the
importance of the order of limits, $N \to \infty$ first and $\e
\to 0$ second.  Physically this means that the frequency resonance is
broad enough to cover a great many modes. Some authors, e.g. ZS and BP
leave the sum notation in the PDF equation even after the $\e \to 0$
limit taken giving $\delta(\omega_{jm}^{n})$.  One has to be careful
interpreting such a formula because formally the RHS is null in most of
the cases because there may be no exact resonances between the
discrete $k$ modes (as it is the case, e.g. for the capillary
waves). Thus, our functional integral notation is a more accurate way
to write the result.

\section{In what sense are the amplitudes independent?}

Obviously, the variables $s_j$ do not separate in the above equation
for the PDF.  Substituting
\BE
{\cal P}^{(N,a)} =
 P^{(a)}_{j_1}  P^{(a)}_{j_2} \dots  P^{(a)}_{j_N} \; 
\label{pure}
\EE
(compare with (\ref{second})) into the discrete version of
(\ref{flux}) we see that it turns into zero on the thermodynamic
solution with $P^{(a)}_{j} = \omega_j \exp(-\omega_j s_j)$.  However,
it is not zero for the one-mode PDF $P^{(a)}_{j}$ corresponding to the
cascade-type Kolmogorov-Zakharov (KZ) spectrum $n_j^{kz}$, i.e.
$P^{(a)}_{j} = (1/n_j^{kz}) \exp(-s_j /n_j^{kz})$ (see next section),
nor it is likely to be zero for any other PDF of form (\ref{pure}).
This means that, even if initially independent, the amplitudes will
correlate with each other at the nonlinear time. Does this mean that
the existing WT theory, and in particular the kinetic equation, is
invalid?

To answer to this question let us differentiate the discrete version
of the equation (\ref{Zequat}) with respect to $\lambda$'s to get
equations for the amplitude moments. We can easily see that
\BE
\partial_t \left(\langle A_{j_1}^2 A_{j_2}^2 \rangle 
-\langle A_{j_1}^2 \rangle  \langle A_{j_2}^2 \rangle \right) =
O(\e^4) \quad (j_1, j_2 \in {\cal B}_N) 
\label{split}
\EE  
if $\langle A_{j_1}^2 A_{j_2}^2 A_{j_3}^2 \rangle = \langle A_{j_1}^2
\rangle \langle A_{j_2}^2 \rangle \langle A_{j_3}^2 \rangle $ (with
the same accuracy) at $t=0$.  Similarly, in terms of PDF's
\BE
\partial_t \left(P^{(2,a)}_{j_1, j_2} (s_{j_1}, s_{j_2}) 
- P^{(a)}_{j_1}(s_{j_1}) P^{(a)}_{j_2}(s_{j_2})  \right) =
 O(\e^4) \quad (j_1, j_2  \in {\cal B}_N)
\EE  
if $P^{(4,a)}_{j_1, j_2, j_3, j_4} (s_{j_1}, s_{j_2}, s_{j_3},
s_{j_4}) = P^{(a)}_{j_1}(s_{j_1}) P^{(a)}_{j_2}(s_{j_2})
P^{(a)}_{j_3}(s_{j_3}) P^{(a)}_{j_4}(s_{j_4}) $ at $t=0$.  Here
$P^{(4,a)}_{j_1, j_2, j_3, j_4} (s_{j_1}, s_{j_2}, s_{j_3}, s_{j_4})
$, $P^{(2,a)}_{j_1, j_2} (s_{j_1}, s_{j_2}) $ and $P^{(a)}_{j} (s_{j})
$ are the four-particle, two-particle and one-particle PDF's obtained
from $\cal P$ by integrating out all but 4,2 or 1 arguments
respectively.  One can see that, with accuracy $\e^2$, the Fourier
modes will remain independent of each other in any pair over the
nonlinear time if they were independent in every triplet at $t=0$.

Similarly, one can show that the modes will remain independent over
the nonlinear time in any subset of $M<N$ modes with accuracy $M/N$
(and $\e^2$) if they were initially independent in every subset of
size $M+1$. Namely
\BEA
P^{(M,a)}_{j_1, j_2, \dots , j_M} (s_{j_1}, s_{j_2}, s_{j_M}) 
- P^{(a)}_{j_1}(s_{j_1}) P^{(a)}_{j_2}(s_{j_2} ) \dots
P^{(a)}_{j_M}(s_{j_M} )   =
O(M/N) + O(\e^2) \nonumber \\
\quad (j_1, j_2, \dots, j_M \in {\cal B}_N)
\EEA  
if $P^{(M+1,a)}_{j_1, j_2, \dots, j_{M+1}} = P^{(a)}_{j_1}
P^{(a)}_{j_2} \dots P^{(a)}_{j_{M+1}} $ at $t=0$.

The mismatch $O(M/N)$ arises from some terms in the ZS equation with
coinciding indices $j$. For $M=2$ there is only one such term in the
$N$-sum and, therefore, the corresponding error is $O(1/N)$ which is
much less than $O(\epsilon^2)$ (due to the order of the limits in $N$
and $\epsilon$).  However, the number of such terms grows as $M$ and
the error accumulates to $O( M/N)$ which can greatly exceed
$O(\epsilon^2)$ for sufficiently large $M$.

We see that the accuracy with which the modes remain independent in a
subset is worse for larger subsets and that the independence property
is completely lost for subsets approaching in size the entire set, $M
\sim N$.  One should not worry too much about this loss because $N$ is
the biggest parameter in the problem (size of the box) and the modes
will be independent in all $M$-subsets no matter how large.  Thus, the
statistical objects involving any {\em finite} number of particles are
factorisable as products of the one-particle objects and, therefore,
the WT theory reduces to considering the one-particle objects.  This
results explains why we re-defined RPA in its relaxed ``essential
RPA'' form.  Indeed, in this form RPA is sufficient for the WT closure
and, on the other hand, it remains valid over the nonlinear time. In
particular, only property (\ref{split}) is needed, as far as the
amplitude statistics is concerned, for deriving the 3-wave kinetic
equation, and this fact validates this equation and all of its
solutions, including the KZ spectrum which plays an important role in
WT.

The situation where modes can be considered as independent when taken
in relatively small sets but should be treated as dependent in the
context of much larger sets is not so unusual in physics. Consider for
example a distribution of electrons and ions in plasma.  The full
$N$-particle distribution function in this case satisfies the Louville
equation which is, in general, not a separable equation. In other
words, the $N$-particle distribution function cannot be written as a
product of $N$ one-particle distribution functions. However, an
$M$-particle distribution can indeed be represented as a product of
$M$ one-particle distributions if $M \ll N_D$ where $N_D$ is the
number of particles in the Debye sphere. We see an interesting
transition from a an individual to collective behavior when the
number of particles approaches $N_D$. In the special case of the
one-particle function we have here the famous mean-field Vlasov
equation which is valid up to $O(1/N_D)$ corrections (representing
particle collisions).

\section{One-particle statistics}

We have established above that 
the one-point statistics are at the heart of WT theory.
All one-point statistical objects can be derived from the one-point
amplitude generating function, 
$$Z_a (\lambda_j) = \left< e^{\lambda_j A_j^2} \right> $$ which can be
obtained from the $N$-point $Z$ by taking all $\mu$'s and all
$\lambda$'s, except for $\lambda_j$, equal to zero.  Substituting such
values into (\ref{Zequat}) we get the following equation for $Z_a$,
\begin{equation}
\frac{\partial Z_a}{\partial t} = \lambda_j \eta_j Z_a +(\lambda_j^2
\eta_j - \lambda_j \gamma_j) \frac{\partial Z_a}{\partial \lambda_j},
\label{za}
\end{equation}
where,
\BEA \eta_j  = 4 \pi \epsilon^2 \int 
\left(|V^j_{lm}|^2 \delta^j_{lm}  \delta(\omega^j_{lm})
+2 |V^m_{jl}|^2 \delta^m_{jl}  \delta(\omega^m_{jl} )
\right)  n_{l} n_{m}
\, d { k_l} d { k_m} ,  \label{RHO} \\
 \gamma_j =
8 \pi \epsilon^2 \int  
\left(
|V^j_{lm}|^2 \delta^j_{lm} \delta(\omega^j_{lm}) n_{m}
 +|V^m_{jl} |^2 \delta^m_{jl} \delta(\omega^m_{jl}) (n_{l}- n_{m})
\right) \, d { k_l} d { k_m}  .  \label{GAMMA}
\EEA
Correspondingly, for the one particle PDF $P_a (s_j) $ we have
\begin{equation}
{\partial P_a \over \partial t}+  {\partial F \over \partial s_j}  =0,
 \label{pa}
\end{equation}
 with $F$ is a probability flux in the s-space,
\begin{equation}
F=-s_j (\gamma P_a +\eta_j {\delta P_a \over \delta s_j}).
\label{flux1}
\end{equation}
Equations (\ref{za}) and (\ref{pa}) where previously obtained and
studied in \cite{cln} in for  four-wave systems.  The only
difference for the four-wave case was different expressions for $\eta$
and $\gamma$. For the three-wave case, the equation for the PDF was not
considered before, but equations for its moments were derived and
solved in \cite{ln}.  In particular, the equation for the first moment is
nothing but the familiar kinetic equation $\dot n = - \gamma n + \eta$
which gives $\eta = \gamma n$ for any steady state. This, in turn
means that in the steady state with $F=0$ we have $P^{(a)}_{j} =
(1/n_j) \exp(-s_j /n_j)$ where $n_j$ can be any steady state solution
of th kinetic equation including the KZ spectrum which plays the
central role in WT \cite{Zakfil,ZLF}.  However, it was shown in
\cite{cln} that there also exist solutions with $F\ne 0$ which
describe WT intermittency.

\section{ Discussion }

In the present paper, we considered the evolution of the full N-particle
objects such as the generating functional and the probability density
function for all the wave amplitudes and their phase factors. We
proved that the phase factors, being statistically independent and
uniform on $S^1$ initially, remain so over the nonlinear evolution
time.  This result does not rely on any assumptions about the
amplitude statistics and, therefore, can be used in future for
studying systems with correlated amplitudes (but random phases).  If
in addition the initial amplitudes are independent too, then they
remain so over the nonlinear time in a coarse-grained sense.  Namely,
all joint PDF's for the number of modes $M \ll N$ split into products
of the one-particle densities with $O(M/N)$ accuracy. Thus, the full
$N$-particle PDF does not get factorized as a product of $N$
one-particle densities and the Fourier modes in the set considered as
a whole are not independent. However, the wave turbulence closure only
deals with the joint objects of the finite size $M$ of variables while
taking the $N \to \infty$ limit. These objects do get factorized into
products and, for the WT purposes, the Fourier modes can be
interpreted as statistically independent.  These results reduce the WT
problem to the study of the one-particle amplitude PDF's and they
validate the generalized RPA technique introduced in
\cite{ln,cln}. Such a study of the one-particle PDF and the high-order
momenta of the wave amplitudes was done in \cite{ln,cln} and the
reader is referred to these papers for the discussion of  WT
intermittency.

Finally, we would like to mention the role of quasi-resonant
interactions which, as we saw, do not produce any long-term effect at
the $\epsilon^2$ order considered in this paper. However, these
interactions do modify statistics at $\epsilon^4$ order as  was
shown in \cite{jansen}.  The $\epsilon^4$ correction can be important
for the real space correlators which have Gaussian values at the
$\epsilon^2$ order for any (not necessarily Rayleigh) amplitude
distributions.


\section{Acknowledgement}

We thank Alan Newell for the feedback he gave us about our results
during his RPI visit in March 2004, particularly for pointing out that
a very similar PDF equation was in derived in Prigogine's book. 


We also thank Colm Connaughton for his comments and for 
the proofreading of the manuscript. 

Yeontaek Choi's work is supported by KOSEF
M07-2003-000-10003-0.  Yuri Lvov acknowledges support provided by NSF
CAREER grant DMS 0134955 and by ONR YIP grant N000140210528.


\end{document}